\documentclass[fleqn,10pt]{wlscirep}
\usepackage[utf8]{inputenc}
\usepackage[T1]{fontenc}
\usepackage{mathtools}
\usepackage{amsmath}
\usepackage{etoolbox}

\usepackage{hyperref}
\usepackage{array}
\usepackage{siunitx}
\usepackage[user,titleref]{zref}
\newcolumntype{M}[1]{>{\centering\arraybackslash}m{#1}}

\usepackage[normalem]{ulem}

\newcommand{\tabref}[1]{Table~\ref{tab:#1}}
\newcommand{\figref}[1]{Fig.~\ref{fig:#1}}
\newcommand{\subfigref}[2]{Fig.~\hyperref[fig:#1]{\ref*{fig:#1}#2)}}
\newcommand{\subfigrefL}[2]{Figure~\hyperref[fig:#1]{\ref*{fig:#1}#2)}}
\newcommand{\subfigsref}[3]{Figs.~\hyperref[fig:#1]{\ref*{fig:#1}#2)}-\hyperref[fig:#1]{\ref*{fig:#1}#3)}}

\definecolor{green2}{rgb}{0.0, 0.5, 0.0}

\title{Optical properties and dynamics of direct and spatially and momentum indirect excitons in AlGaAs/AlAs quantum wells}

\author[1,*]{Dabrówka Biega\'{n}ska}
\author[1]{Maciej Pieczarka}
\author[1]{Krzysztof Ryczko}
\author[1]{Maciej Kubisa}
\author[2]{Sebastian Klembt}
\author[2]{Sven H\"{o}fling}
\author[3]{Christian Schneider}
\author[1,+]{Marcin Syperek}

\affil[1]{Department of Experimental Physics, Faculty of Fundamental Problems of Technology, Wroc\l{}aw University of Science and Technology, Wybrzeże Wyspia\'{n}skiego 27, 50-370 Wroc\l{}aw, Poland}
\affil[2]{Julius-Maximilians-Universit\"{a}t W\"{u}rzburg, Physikalisches Institut and W\"{u}rzburg-Dresden Cluster of Excellence ct.qmat, Lehrstuhl f\"{u}r Technische Physik, Am Hubland, 97074 W\"{u}rzburg, Germany}
\affil[3]{Institute of Physics, University of Oldenburg, D-26129 Oldenburg, Germany}
\affil[*]{dabrowka.bieganska@pwr.edu.pl}
\affil[+]{marcin.syperek@pwr.edu.pl}

\begin{abstract}
{We present an experimental study on optical properties and dynamics of direct and spatially and momentum indirect excitons in AlGaAs/AlAs quantum wells near the crossover between $\varGamma-$ and $X$-valley confined electron states. The time-integrated photoluminescence experiment at $T=$ \SI{4.8}{\kelvin} revealed three simultaneously observed optical transitions resulting from (a) a direct exciton recombination, involving an electron and a hole states both located in the $\varGamma$-valley in the quantum well layer, and (b) two spatially and momentum indirect excitons, comprising of the confined electron states in the $X$-valley in the AlAs barrier with different effective masses and quantum well holes in the $\varGamma$-valley. This interpretation has been based on the optical pumping density-dependent, temperature-dependent and spatially-resolved photoluminescence measurements, which provided the characterization of the structure, crucial in potential system's applications. Additionally, the time-resolved photoluminescence experiments unveiled complex carrier relaxation dynamics in the investigated quantum well system, which is strongly governed by a non-radiative carrier recombination - the characteristics further
critical in potential system’s use. This solid state platform hosting both direct and indirect excitons in a highly tunable monolithic system can benefit and underline the operation principles of novel
electronic and photonic devices.
}    
\end{abstract}
\begin{document}

\flushbottom
\maketitle

\section*{Introduction}

A discovery and recent interest in two-dimensional (2D) materials spark remarkable progress in research on excitons and the use of excitonic properties in optoelectronic devices. Particular attention has been paid to spatially indirect excitons with Coulomb-bound electrons and holes characterized by spatially separated wavefunctions. The strong dipolar nature of such excitations allows for the observation of new exciting phenomena, including a dipolar excitonic insulator~\cite{Gu2022}, high-temperature superfluidity~\cite{Fogler2014}, and high-temperature Bose-Einstein condensation~\cite{Butov1994,Butov2002,Wang2019}, among others. Moreover, due to their high dipole moment, prolonged lifetime, high diffusion length and bosonic nature, as well as good control of the exciton properties due to their sensitivity to an electric field and dielectric environment, indirect excitons can benefit and underline the operation principles of novel electronic and photonic devices~\cite{Ciarrocchi2022,Yulun2024}. In this new wave of exciting discoveries, the attention can be brought back to the relatively mature AlGaAs/AlAs quantum well (QW) system in which similar indirect excitons can be generated in a highly controllable environment. Surprisingly, the knowledge on the isolated AlGaAs/AlAs QW system and its excitonic properties is limited in the literature  \cite{Masumoto1988,Dawson1989,Young1990,Lee1996,Haetty1997}, with research focused mainly on superlattice systems, not on isolated QWs.

What is important is that this system offers fabrication capabilities that are still unreachable in 2D layered semiconductors. The use of III-V semiconductors, in particular with AlGaAs alloy materials in monolithically integrated structures is highly beneficial due to the high quality of epitaxial growth, precise control over the structure parameters, and small lattice mismatches necessary to achieve high quality factors of optical microcavities. They can be easily integrated into multilayer monolithic structures and are suitable for a wide range of well-developed microelectronic processing techniques. On the other hand, the important drawback in developing useful systems and studying new excitonic-based physical phenomena in these materials is a low exciton binding energy, particularly when compared with 2D materials such as transition-metal dichalcogenides or 2D perovskites. The typical energies on the order of a few millielectronvolt are too small compared to the thermal energy at increased temperatures, which results in the exciton's effective dissociation, preventing an elevated-temperature device operation. 

Nevertheless, several attempts have been made to overcome this issue. One of the approaches employs the $\varGamma-X$ band mixing  \cite{Ting1987,Ru2003,Dawson1989}. As GaAs is a direct bandgap semiconductor at the $\varGamma$ point, while AlAs is characterised by an indirect bandgap with the global minimum of the conduction band in the $X$ point of the Brillouin zone, the Al$_x$Ga$_{1-x}$As alloy is characterized by a direct-to-indirect bandgap transition at a specific proportion of the two binary semiconductors of around $x = 0.24$  \cite{Fluegel_2015,Chand1984}. Near the $\varGamma-X$ crossover, the quantum mechanical mixing of the $\varGamma-$ and the $X-$valley electrons was reported to manifest in large enhancement of the donor activation energies and the exciton binding energies  \cite{Pearah1985}. At the same time, the nearly resonant spatially separated levels open a possibility for indirect excitons to emerge.
 
 Near-resonant levels of $\varGamma$ QW electrons and $X$ AlGaAs states were studied in single and multi-QW structures  \cite{Tada1988,Feldmann1992,VanKesteren1989}, as well as in superlattices  \cite{Ting1987,Ru2003,Birkedal1997,Cingolani1989,Kato1989,Peterson1988,Danan1987}, showing the possibility of tunnelling  \cite{Feldmann1992} and interactions  \cite{Ru2003} between the confined electron states. Such interacting states have yet to be studied in more complex structures, such as enclosed microcavities, where complex dynamics seems crucial in the structure design and band structure engineering.

 In this report, we bring the III-V-based system back to the spotlight, uncovering the properties that can underline the operation of exciton-based photonic devices. We studied the optical properties of AlGaAs/AlAs QWs, operating close to the $\varGamma-X$ coupling regime.  Using photoluminescence experiments supported by theoretical calculations within the effective-mass framework, we unravel the nature of multiple resonances present in the structure spectra and interpret two of the lowest energy states as spatially- and momentum-indirect X-excitons. We characterize the density and temperature evolution of all the energy states, as well as their temporal decay and spatial diffusion. Our work points to the complex exciton and carrier dynamics in this system, hindered by nonradiative processes and further affected by localization effects. Such detailed investigation on the interplay between the energy states is important to understand the resulting complex dynamics in the full structure such as microcavity, under the additional interaction with photonic modes, particularly of the dipolar excitons. Investigating the system in a wide range of temperatures is necessary to carefully design and understand future high-temperature devices.

\section*{Results}

\subsection*{Steady-state photoluminescence spectroscopy}

\begin{figure*}[h]
    \centering
    \includegraphics[width=15cm]{figure1.pdf}
    \caption{\label{fig:fig1}}{a) Schematic illustration of the investigated quantum well (QW) structure. b)  Low temperature ($T=4.8\ K$) photoluminescence (PL) spectra of the Al$_{0.20}$Ga$_{0.80}$As /AlAs QWs registered at quasi-resonant excitation, $E_{exc}=$ \SI{2.00}{\electronvolt}, and low excitation power $P_{exc}\approx$\SI{10}{\micro\watt}. c) Scheme of the band structure of one period of the structure with the single-particle levels marked with dashed lines the and optical transitions indicated with arrows. The solid black line represents the edge of the valence band (VB), while solid grey and green lines show the edges of the conduction band (CB) in the $\varGamma$ and $X$ valley, respectively.}
  
\end{figure*}

The structure under study is schematically depicted in \subfigref{fig1}{a}. It consists of twelve \SI{9}{\nano\meter}-wide Al$_{0.20}$Ga$_{0.80}$As QWs, separated by \SI{4}{\nano\meter} AlAs barriers, distributed in three stacks of four, comprising the active part of the structure. Good isolation of each QW, provided by the barriers, ensures negligible coupling of confined states between neighbouring QWs. The active part was monolithically integrated on a distributed Bragg reflector (DBR), enhancing emission involving confined QW states. More details on the sample can be found in the \textbf{\ztitleref{Methods}} section.

To probe the fundamental excitations in this QW system, we employ the steady-state photoluminescence (PL) experiment (see~\textbf{\ztitleref{Methods}}). \subfigrefL{fig1}{b} displays the measured low-temperature ($T= \SI{4.8}{\kelvin}$) PL spectrum of the AlGaAs/AlAs QW, registered under a quasi-resonant energy excitation at $E_{exc}= \SI{2.00}{\electronvolt}$ and an excitation power of $P_{exc}=\SI{10}{\micro\watt}$. The PL spectrum consists of three well-resolved features, with enhanced visibility due to the DBR. The features originate from the carrier recombination between QW-confined states. The highest-energy and highest-intensity PL band located at \SI{1.8472}{\electronvolt} corresponds well to the Coulomb-correlated electron-hole (exciton) recombination in the vicinity of the $\varGamma$ valley in the Brillouin zone. The transition involves the lowest-lying $\varGamma$-electron state and the topmost $\varGamma$-hole state confined in the AlGaAs QW. The assignment has been suggested by the energy match with the calculations performed during the initial design of the sample \cite{Suchomel2017}, as well as the PL band intensity dominating the spectrum. Therefore, it should reflect the recombination of momentum- and spatially direct $\varGamma$-exciton (annotated simply as $\Gamma$), having a high optical transition probability.

The presence of two PL bands located energetically below the $\Gamma$ exciton is initially surprising. They are centred at energies of \SI{1.8007}{\electronvolt} and \SI{1.7843}{\electronvolt} respectively, redshifted by nearly \SI{47}{\milli\electronvolt} and \SI{63}{\milli\electronvolt} from the direct $\Gamma$ exciton transition. Large energy separation with respect to $\Gamma$ excludes their identification as charged-excitons or multi-exciton recombination since these are typically characterized by sub-\SI{10}{\milli\electronvolt} binding energies~\cite{Manassen1996ExcitonWells,Syperek2007SpinWells}. The defect-assisted recombination would likely appear in a spectrum as a single broad transition instead of two clearly resolved bands, resembling the excitonic transitions in the AlGaAs/AlAs QW~\cite{Haverkort_1992}. Therefore, we attribute the observed spectral features to the recombination of excitons involving the $X$-valley electrons and $\varGamma$-valley holes, observed previously in other AlGaAs nanostructures\cite{Lee1996,VanKesteren1989,Young1990}. We further investigate the nature of these transitions in the following sections.

\subsection*{Calculations} \zlabel{calc}
To elucidate the nature of the two optical transitions energetically below the $\Gamma$ exciton and to provide arguments for their identification, the QW band structure has been calculated within the effective mass approximation. \subfigref{fig1}{c} shows the band alignment for Al$_{0.20}$Ga$_{0.80}$As QW material and the AlAs barriers in the conduction (CB) and valence (VB) bands, including the electrons from both the $\varGamma$ and $X$ valleys of the Brillouin zone. It is important to note that while in the $\varGamma$ valley the CB profile (\subfigref{fig1}{c}, solid grey line) creates a confinement potential for electrons in the QW material, the confinement for $X$-valley electrons (\subfigref{fig1}{c}, solid green line) is created in the AlAs barrier \cite{Lee1996,Ihm1987,O.Gobel1990}. Holes can be confined only within the $\varGamma$-valley in the QW material \cite{Lee1996}. For clarity, only the topmost heavy hole ($hh$) band is presented, and unstrained materials are considered, in good agreement with the experimental implementation. 

The band alignment is the basis for calculations of the single-particle confined states for $\varGamma$ electrons ($e_1$) and holes ($hh_1$), and $X$ electrons ($x_{x,y}$, $x_z$), as schematically depicted in \subfigref{fig1}{c}. The presence of two $X$-electron confined states results from the $X$-electron effective mass anisotropy, well known for the AlAs material \cite{Maezawa1992,Im1999,Ting1987}. As the quantization occurs in the growth direction perpendicular to the QW plane (the $z$-direction, [001] in the crystal structure), two $X$-electron masses need to be taken into account in calculations, in contrast to the isotropic $\varGamma$ conduction band  \cite{O.Gobel1990,Ting1987}. This results in a doublet of states for transverse, $x_{x,y}$, and longitudinal, $x_z$, electrons in the $X$ valley. Their masses differ by a factor of $0.2268$ in the AlAs layer (with a higher value of the longitudinal mass) \cite{Piperek2003}, yielding the energy separation of approximately \SI{28}{\milli\electronvolt} between the calculated energy levels in the doublet. Our calculations do not include strain effects that can lead to further state separation \cite{Pietka2007}.

$\varGamma$ and $X$ electrons in confined states can interact with $\varGamma$ holes confined in the QW layer via Coulomb interactions, forming spatially and momentum direct and indirect excitons. In our calculations, we estimated the Coulomb corrections to single-particle energies using the Rydberg formula (see \textbf{\ztitleref{Methods}}). This simplified approach is likely to be an overestimation of the actual electron-hole Coulomb coupling for the indirect excitons due to the spatial separation between carriers \cite{Wu2015,Piperek2003}. On the contrary, the binding energy of the $\Gamma$ exciton is underestimated due to the two-dimensional nature of the QW states \cite{O.Gobel1990}. Additionally, the $\varGamma-X$ coupling (neglected in our calculations) may affect the exciton energies \cite{Pearah1985}. More accurate evaluations of the exciton energies require complex self-consistent calculations, which are beyond the scope of this work \cite{Chang1994,Wu2015,PONOMAREV2005539}. Nevertheless, our simplified approach results in good agreement with experimental observations, as discussed below.

The calculated band structure and the energies of the optical transitions are confronted with the results of the PL measurement in \tabref{tab1}. The data shows that the Coulomb-corrected energy difference between the $e_1$ and the $hh_1$ states in the $\varGamma$-band corresponds very well to the spectral position of the most intensive PL band identified as $\Gamma$. As expected, the type-I confinement and the direct nature of the transition in the real and in the momentum space are linked to its high optical intensity. More importantly, the calculated energies of transitions involving the $x_{x,y}$ and $x_z$ states in the $X$-valley and the $\varGamma$-valley heavy hole state (marked as $X_{X,Y}$, and $X_Z$ in \subfigref{fig1}{c}) are in good agreement with the PL band energies observed below the $\Gamma$. Both $X_{X,Y}$, and $X_Z$ transitions are nominally indirect in the real and momentum space, which should lead to a low transition probability. However, several effects can increase this value. It is important to note that as the translational symmetry is broken in the growth direction ($z$), the $X_Z$ recombination is allowed without phonon assistance, regardless of its indirect character in the momentum space, due to the weakening of the momentum selection rules \cite{Finkman1987,Pietka2007,Danan1987,O.Gobel1990}. Translational invariance of the Hamiltonian (hence the momentum selection rules) should not be broken in the plane of the structure, resulting in the need for phonon presence for the $X_{X,Y}$ transition to occur \cite{Finkman1987,Pietka2007,Danan1987,O.Gobel1990}. The excitonic states can therefore be described as pseudo-direct in the case of $X_Z$ and indirect in the case of $X_{X,Y}$. Notably, both remain indirect in the real space.

Additionally, it is known that in the energetic vicinity of the electron states of the $\varGamma$ and $X$ valleys in the Al-rich AlGaAs  semiconductor compounds, the $x_{x,y}$ and $x_z$ states can have an admixture of the $e_1$ states, which increases the transition probability (this effect is not included in our calculations) \cite{Chang1994,Dawson1989,O.Gobel1990,Ihm1987}. What is more, the $X$-valley electrons can be shifted towards the AlGaAs/AlAs heterointerface due to the Coulomb interaction with holes confined in the QW layer, the localization at interface inhomogeneities (caused by chemical content fluctuations), and local electric field or strain fluctuations, which increases the overlap integral with the $\varGamma$ valley hole states \cite{Dawson1989, Feldmann1992,Finkman1987}. Interface inhomogeneities are expected to have the strongest influence on the $x_z$ electron states, as their effective mass is the largest \cite{Ting1987} (effect visualized in our spatial diffusion measurements, described in the next section).

\begin{table}[h]
\centering
\begin{tabular}{||M{2.6cm}|M{2.6cm}|M{2.9cm}|M{2.9cm}|M{2.9cm}||}
\hline
\textbf{Exciton} & \textbf{PL peak energy $(4.8\ K)$} & \textbf{Calculated single-level transition energy ~~~ $E$} & \textbf{Calculated exciton binding energy ~~~ $E_{Ry}$} & \textbf{Calculated exciton recombination energy ~~~~~~~~~ $E-E_{Ry}$} \\
\hline
\textbf{$\Gamma (e_1 - hh_1) $} & 1.8472 eV & 1.8521 eV & 4.8 meV & 1.8474 eV\\
\hline
\textbf{$X_{X,Y} (x_{x,y} - hh_1) $} & 1.8007 eV & 1.8130 eV & 9.5 meV & 1.8062 eV\\
\hline
\textbf{$X_{Z} (x_{z} - hh_1) $} & 1.7843 eV & 1.7847 eV & 6.8 meV & 1.7752 eV \\
\hline
\end{tabular}
\caption{\label{tab:tab1} Measured and calculated energies of the three investigated optical transitions, including the energies calculated from single-particle states ($E$), calculated excitonic correctons ($E_{Ry}$) and the transition energies at $T= \SI{4.8}{\kelvin}$.
}

\end{table}

\subsection*{Excitation power-dependent photoluminescence}
\zlabel{excitation}

\begin{figure*}[h]
    \centering
    \includegraphics[width=\textwidth]{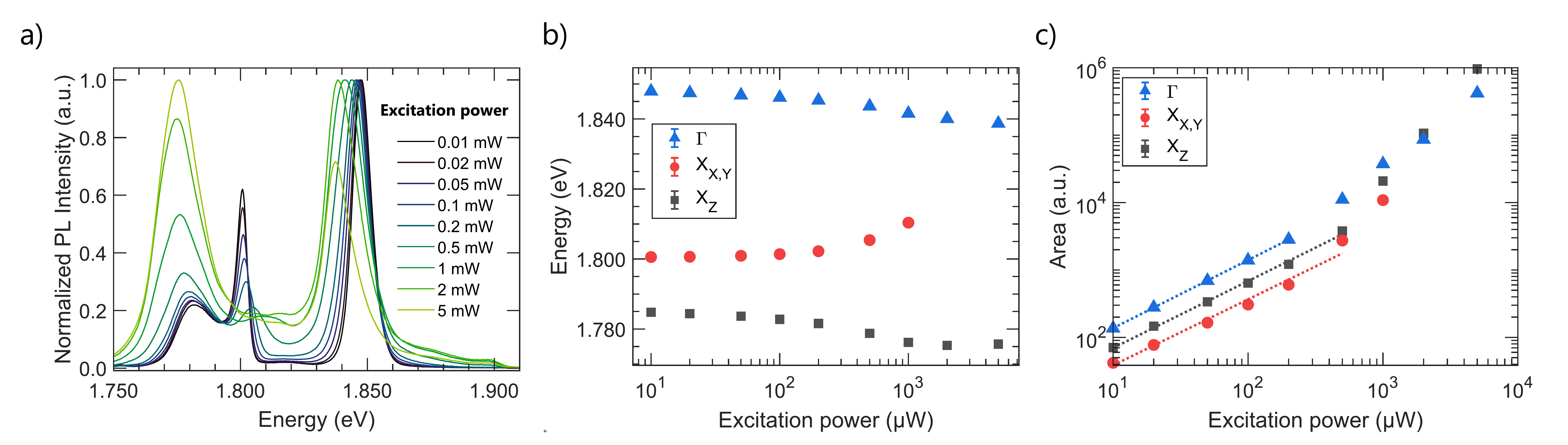}
    \caption{
    a) Excitation power-dependent photoluminescence (PL) spectra at $E_{exc}=\SI{2.00}{\electronvolt}$ and $T=\SI{4.8}{\kelvin}$. 
    b) Change of the PL peak position with excitation power. 
    c) Evolution of the PL intensity (represented as fitted peak area)  of three respective bands as a function of $P_{exc}$. Dashed lines show the power-law fits to the experimental points at low excitation powers (in the linear growth regime), with slopes of $\gamma=$ \num{1.009\pm0.005}, \num{0.997\pm0.005} and \num{0.99\pm0.01} for $\Gamma$, $X_{X,Y}$ and $X_Z$ respectively.
    }
    \label{fig:fig3}
\end{figure*}

The PL emission bands are examined as a function of the excitation power, $P_{exc}$, providing additional information on the nature of observed emission and the underlying optical transitions. \subfigrefL{fig3}{a} presents the spectra evolution with increasing $P_{exc}$. Three PL maxima, corresponding to the $\Gamma$, $X_{X,Y}$, and $X_Z$  excitons, were fitted with Gaussian profiles and the extracted energies are shown in \subfigref{fig3}{b}. One can observe that the middle-energy peak (corresponding to the $X_{X,Y}$ exciton) becomes unresolvable from the spectra under the strongest pumping. In contrast to the other two states, its energy blueshifts as a function of excitation power, what can be attributed to its dipolar nature and the repulsive interactions between quasiparticles, as was observed before for spatially indirect excitons\cite{Lee1996,Butov2001,Butov1999}. Contrarily, both $\Gamma$ and $X_Z$ exciton energies present a small redshift with increasing excitation power, a behaviour not expected for indirect states. It is important to note that in the case of the $X_Z$ excitons, their ground state nature makes them more prone to inhomogeneities, which affect their lifetime and transport properties. Higher probability of the transition due to the weakened momentum selection rules, shorter decay and higher localization can hinder their build-up, necessary for the repulsive interactions to be observed. Small redshift of the $\Gamma$ and the $X_Z$ PL bands at high excitation powers indicate the local sample heating under the elevated pumping.  

\subfigrefL{fig3}{c} shows the spectrally integrated emission intensity of the three investigated bands as a function of $P_{exc}$,  plotted in the log-log scale. Experimental points within the low excitation power range were fitted with the power law function $I\propto P_{exc}^{\gamma}$, where the $\gamma$ value depends on the recombination mechanism  \cite{Schmidt1992,Spindler2019}. 
For the defect-assisted mono-molecular recombination of photo-injected electrons or holes the coefficient is expected to be lower than 1. However, when the recombination is governed by the mixture of free and bound exciton annihilation,  $1<\gamma<2$. Our fitting procedure yielded almost the same $\gamma$ parameter for all three PL bands. Extracted values $\gamma=$ \num{1.009\pm0.005}, \num{0.997\pm0.005} and \num{0.99\pm0.01} for $\Gamma$, $X_{X,Y}$ and $X_Z$ respectively, clearly suggest that all the observed optical transitions have excitonic character in the considered low excitation power range. 

The superlinear growth at increased $P_{exc}$ (see \subfigref{fig3}{c}) is quite surprising. It begins at an excitation power of approximately $\SI{\sim3e2}{\micro\watt}$. At this power level, the calculated electron-hole pair density is estimated as $\SI{\sim3.9e18}{\per\cubic\centi\meter}$ (see \textbf{\ztitleref{Methods}}), what approaches the Auger threshold limit for the QW system, estimated to be on the order of $\SI{\sim e19}{\per\cubic\centi\meter}$  \cite{Wang1990, Borri1997}. Further increase of the excitation power, up to $P_{exc}=\SI{e3}{\micro\watt}$, is expected to hypothetically reach this density limit and lead to a non-radiative recombination taking over the radiative processes, causing a sublinear increase in the PL intensity \cite{Borri1997,Zhu2016,Sumikura2021}. However, the observed superlinear increase contradicts these expectations, suggesting that the number of the photogenerated electron-hole pairs in the QW stack is significantly lower than the estimated value. It could result from the fast, non-radiative carrier relaxation following the pulse photoexcitation, which could efficiently lower the carrier population in the QW. According to this interpretation, a non-radiative state saturation may occur at an excitation density of \SI{\sim3e2}{\micro\watt}, potentially leading to the increased PL intensity form the QW confined states with a further increase of the pumping power. 
These effects can be confirmed by studying the PL temporal decay, as described in section \textbf{\ztitleref{Time-resolved photoluminescence}}.

\subsection*{Temperature-dependent photoluminescence} \zlabel{temp}

\begin{figure*}[h]
    \centering
    \includegraphics[height=6cm]{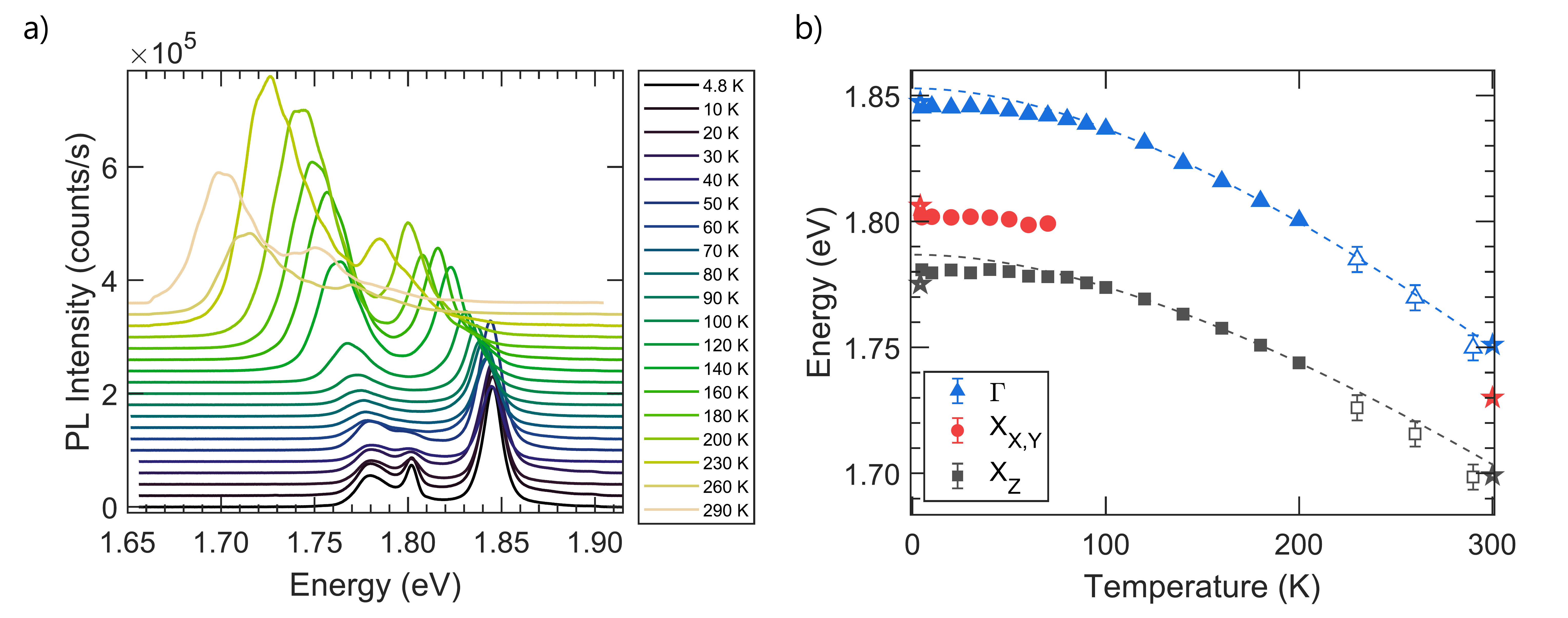}
    \caption{a) Temperature-dependent photoluminescence (PL) spectra at $E_{exc}= \SI{2.00}{\electronvolt}$ and $P_{exc}=\SI{150}{\micro\watt}$. b) Temperature-induced energy shift of the PL bands. Points show the band energies extracted from fitting the spectra in a) with Gaussian profiles, open points mark the energies extracted from the spectra maximal intensity values and stars indicate the excitonic transitions calculated within the effective mass approximation. Dashed curves are the Varshni fits to the temperature dependences above \SI{100}{\kelvin}.}
    \label{fig:fig4}
\end{figure*}

To further characterize the studied system's excitonic features, we look at the temperature evolution of the PL spectra in the wide temperature range, with results presented in \figref{fig4}. As the temperature $T$ increases, the energy of all resonances redshifts, as seen in \subfigsref{fig4}{a}{b}. At $T\approx \SI{80}{\kelvin}$, the $X_{X,Y}$ transition becomes unresolvable, while the neighbouring $X_Z$ starts to dominate the spectrum. We highlight the fact that spectra presented in \subfigref{fig4}{a} are not normalized, and the intensity of $X_Z$ emission strongly increases at $T>\SI{100}{\kelvin}$ reaching the maximum at \SI{\sim200}{\kelvin}. We hypothesize that the intensity increase originates from the rising phonon density, making both the radiative transition from the $X_Z$ exciton and the nonradiative transitions between $e_1$,  $x_{x,y}$, and $x_z$ electron levels more probable. Relaxation of electrons to the $x_z$ state can be additionally increased by the decrease in energy separation between the $\Gamma$ and $X$ states with temperature (due to the different temperature variation of the $\varGamma$ and $X$ energy gaps in AlGaAs and AlAs layers respectively \cite{Vurgaftman2001,Danan1987}), as well as by the quench of the $X_{X,Y}$ emission, making the inter-level transfer more effective. Additionally, these may be contributed to by an increased carrier density in the QW, owing to the temperature-driven release of carriers from nonradiative charge traps. The results of the excitation power-dependent and temperature-dependent time-resolved PL measurements suggest the existence of these traps (see sections  \textbf{\ztitleref{excitation}} and \textbf{Time-resolved photoluminescence}). However, we also note, that absorption and effectiveness of the excitation strongly differ with temperature as the energy gaps shift, with $E_{exc}$  crossing the band gap energy of the Al$_{0.40}$Ga$_{0.60}$As spacer in the structure at  $T\approx \SI{220}{\kelvin}$, while the pumping energy in our experiment was set constant ($E_{exc}=\SI{2.00}{\electronvolt}$). Hence, no definite conclusions based on the intensity evolution can be made. We note that the efficient high-temperature PL highlights the potential use of the structure in photonic devices at elevated temperatures, possibly up to room temperature. 

Additionally, temperature dependences of the transition energies were fitted using the Varshni formula: \cite{VARSHNI1967}\\ 
$E_{\mathrm{t}}(T) = E_{\mathrm{t}}(0)-\alpha T^2/ \left(T+\beta\right)$ 
where $E_{t}$ is the transition energy, and $\alpha$ and $\beta$ are fitting constants related to the used semiconductor materials.
The fitting results are presented in \subfigref{fig4}{b} with dashed lines. Extracted $\alpha$ and $\beta$ constants ($\alpha=$ \SI{7.97e-4}{\electronvolt\per\kelvin\tothe{2}} and $\beta=\SI{399}{\kelvin}$ for the $\Gamma$ state, and the $\alpha=\SI{6.71e-4}{\electronvolt\per\kelvin\tothe{2}}$ and $\beta= \SI{425}{\kelvin}$ for the $X_Z$ state) match well the reported material parameters of GaAs and AlAs \cite{Vurgaftman2001,Lourenco2001}. Curves were fitted to the experimental points above \SI{100}{\kelvin}, as at low temperatures the dependences deviate from the Varshni model. Transition energies lower than the theoretical values point to the likely additional localization effect and Coulomb correlations. At low temperatures, both $\Gamma$ and $X_Z$ excitons can be trapped, e.g. by spatial inhomogeneities of the QW interfaces or the chemical composition fluctuations, acting as localization centres, when the thermal energy is insufficient to overcome the energy potential minima \cite{VanKesteren1989,Colocci_1990}. 
Localisation effect is also affecting the diffusion of the excitonic complexes, showing the limited spatial extent of the low-temperature PL emission of both $\Gamma$ and $X_Z$ states, contrasted with the increase of the $X_Z$'s diffusion at elevated temperatures, as presented and described in section \textbf{\ztitleref{diffusion}}, below.
Localization energies suggested by the difference between the Varshni curves and the measured exciton emission at low temperatures are around \SIrange{6}{8}{\milli\electronvolt}.  

Finally, we estimate the expected exciton recombination energies at room temperature, using the same methods as described in the \textbf{\ztitleref{calc}} section. In contrast to previous estimations \cite{Suchomel2017}, we verify the system to be indirect (with the type-II transition from the quantized X-valley state in the barrier to the first heavy hole QW energy state as the lowest energy transition) in the whole studied temperature range up to the room temperature. At \SI{300}{\kelvin}, the exciton energies estimated from our calculations are \SI{1.751}{\electronvolt}, \SI{1.730}{\electronvolt} and \SI{1.6992}{\electronvolt} for the $\Gamma$, $X_{X,Y}$ and $X_Z$ transitions respectively, matching nearly perfectly the PL resonances observed in the experiment (see \subfigref{fig4}{b}, open stars).

\subsection*{Exciton diffusion} \zlabel{diffusion}
\begin{figure*}[h]
    \centering
    \includegraphics[height=5.5cm]{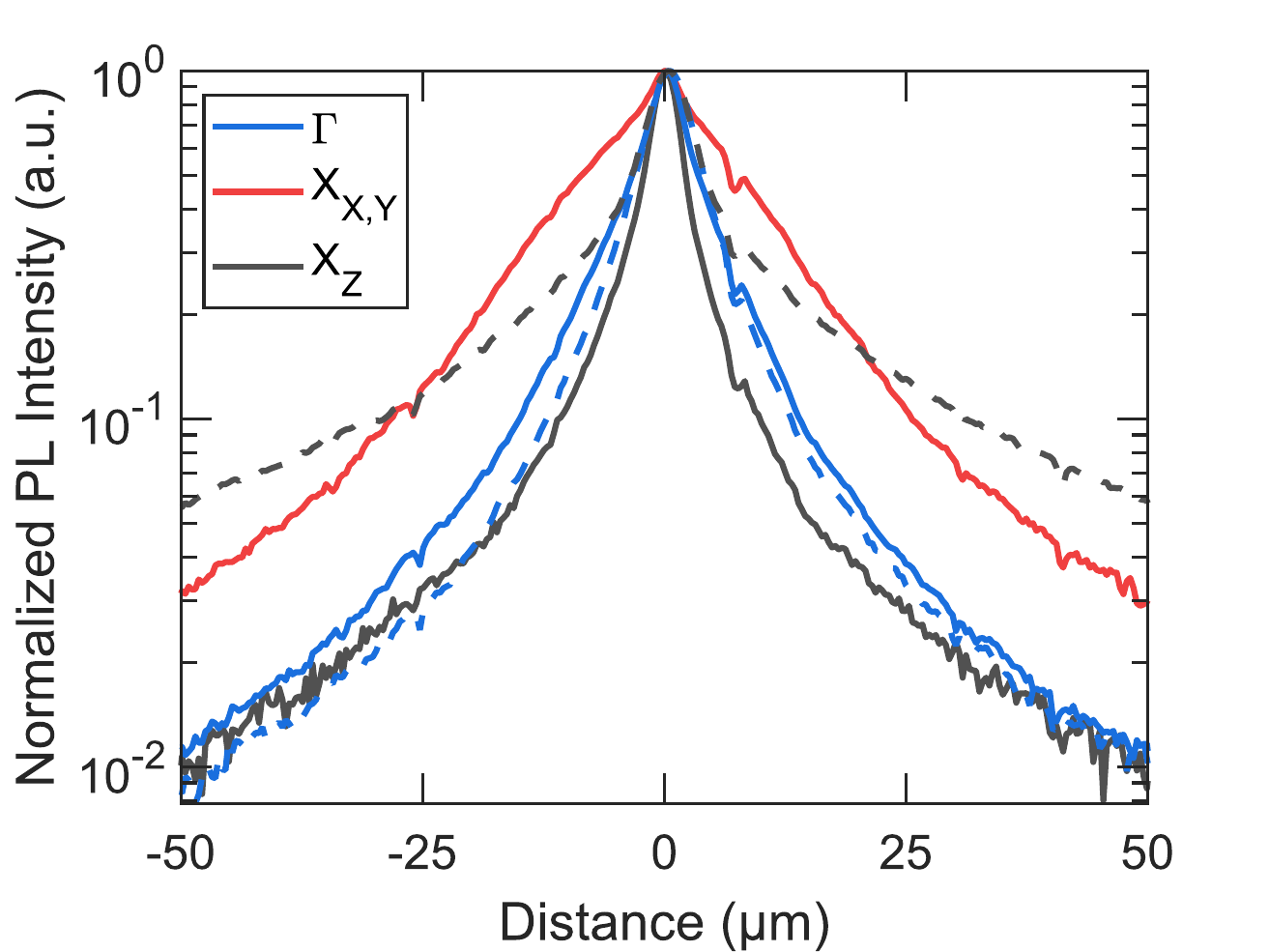}
    \caption{Spatial extent of the photoluminescence intensity, showing diffusion profiles of the three studied excitons, measured at low excitation power ($P_{exc}=\SI{10}{\micro\watt}$) and the temperature of $T=\SI{4.8}{\kelvin}$ (solid lines) and $T=\SI{140}{\kelvin}$ (dashed lines). The $X_{X,Y}$ exciton is not visible at elevated temperatures.}
    \label{fig:fig2}
\end{figure*}

The assignment of the observed QW transitions is further supported by the measured spatial profiles of the time-integrated PL emission at low temperatures and low excitation power density. The spatial extent of the PL signal of the three observed transitions at $T=\SI{4.8}{\kelvin}$ is displayed in \figref{fig2} (solid lines). One can see a clear difference between the emission profiles of the studied states. The $X_{X,Y}$ excitons (the middle-energy-peak in \subfigref{fig1}{b}) travel further away from the excitation spot than the $\Gamma$ excitons before recombining, with their emission profile extending for more than \SI{100}{\micro\meter}. The $X_Z$ excitons are characterized by the smallest in-plane diffusion profile at low temperatures, even slightly smaller than that of the $\Gamma$ excitons. Extended diffusion profiles are commonly observed for indirect excitons, having high momentum, and often being characterized by long radiative lifetimes, thus propagating macroscopic distances before recombination  \cite{Ivanov_2002,Butov2002,Fasol1989}. Several factors can influence the diffusion, including the in-plane effective mass, phonon density, exciton-phonon interaction and spatial localization of the states, as well as the efficiency of a non-radiative recombination \cite{Heller1996, Smith1989,Ivanov_2002}. The need for the phonon assistance in the $X_{X,Y}$ transition affects its recombination probability, hence the diffusion length, explaining its stark distinction from the other two profiles. In the case of the $X_Z$ excitons, the aforementioned breaking of the momentum-selection rules makes them more similar to the $\Gamma$ excitons, what is visualized in the narrower diffusion profile. Moreover, having the largest effective mass in the z-direction, their transport properties might be influenced by the spatial inhomogeneities of the QW interfaces, which act as localization centres  \cite{Butov2002,Butov1998}, further affecting the diffusion. 

The localization effect is clearly visualized by studying the diffusion profiles at elevated temperatures (experimental curves measured at $T=\SI{140}{\kelvin}$ shown in \figref{fig2} with dashed lines). The diffusion of the $X_Z$ state largely increases at elevated temperature and becomes significantly broader spatially than that of the $\Gamma$ state, which remains narrow. The observation can be linked to the nature of an indirect state and the localization. Once the thermal activation overcomes the localization energy from local potential minima, the diffusion range of the $X_Z$ exciton is expected to be longer than that of the direct excitons. On the other hand, it remains narrower than the $X_{X,Y}$ transition at low temperatures due to the difference in selection rules and in the effective mass. Full characterization of the temperature dependence is discussed in 
the previous section, where more localization effects become apparent. 
Overall, observed differences in the diffusion profiles support our interpretation of the nature of the direct and indirect excitons, as similar behaviour has been observed before in this material system \cite{Pietka2007}. 

\subsection*{Time-resolved photoluminescence}
\zlabel{Time-resolved photoluminescence}

\begin{figure*}[h]
    \centering
    \includegraphics[height=7cm]{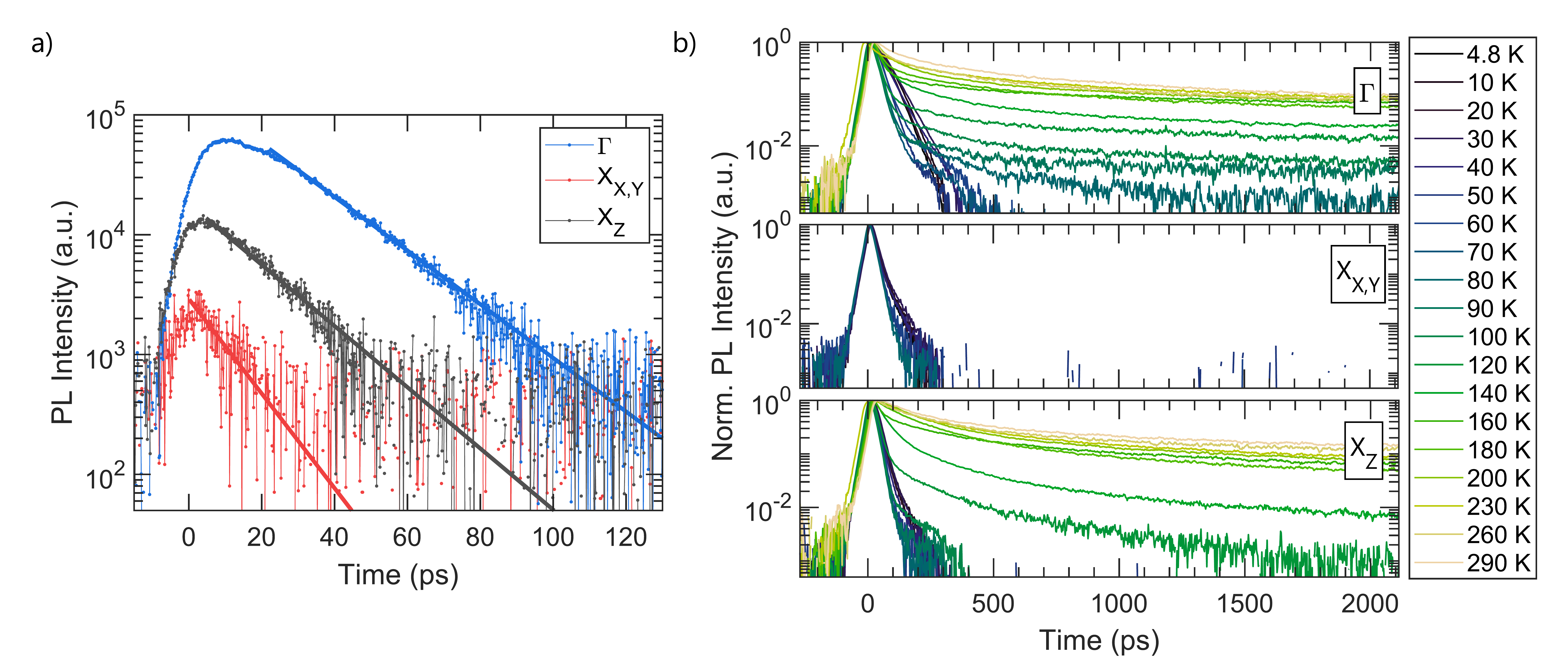}
    \caption{a) Temporal evolution of the photoluminescence (PL) intensity of the three studied transitions recorded at low temperature $T=\SI{4.8}{\kelvin}$ and excitation power $P_{exc}=\SI{100}{\micro\watt}$. Solid lines show monoexponential decay curves, fitted to experimental data (points), with the time constants of \SI{11}{\pico\second}, \SI{17}{\pico\second} and \SI{19}{\pico\second} for $X_{X,Y}$, $X_Z$ and $\Gamma$ exciton respectively. b) Temporal PL decays of three investigated transitions with increasing temperature. Excitation energy ($E_{exc}=\SI{2.00}{\electronvolt}$) and power ($P_{exc}=\SI{150}{\micro\watt}$) were set constant throughout the series.}
    \label{fig:fig5}
\end{figure*}

Finally, we explore the exciton recombination dynamics in the system. \subfigrefL{fig5}{a} shows the temporal evolution of the PL intensity for the $\Gamma$, $X_{X,Y}$ and $X_Z$ transitions, following a \SI{\sim200}{\femto\second} excitation pulse at low temperature ($T=\SI{4.8}{\kelvin}$) and at the excitation power of $P_{exc}=\SI{100}{\micro\watt}$. 

All three transitions can be approximated by a mono-exponential decay with the characteristic decay time constant ($\tau_{PL}$) of about \SI{11}{\pico\second} for the $X_{X,Y}$ transition and \SI{17}{\pico\second} for the $X_Z$. We note that such short decays are close to the temporal resolution of our detection system (of \SI{\sim7}{\pico\second}), and are much smaller than the decay times expected for direct\cite{Syperek2007SpinWells,Vinattieri1994,Deveaud1991,Citrin1993} and indirect transitions \cite{Langbein1996,Dawson1989,Baranowski2011,Finkman1987,Feldmann1992,Harpaz2017}. This observation reveals the likely presence of efficient nonradiative decay channels for the photoinjected carrier population in our system, dominating the measured dynamics and manifested by a shortening of the PL decay. 

Among the processes that could lead to the shortening of the PL decay time at cryogenic temperatures is an efficient depopulation of the QW reservoir through carrier relaxation into the states located below the fundamental QW bandgap, or the Auger-like processes, expected mainly in the high pumping regime. Importantly, the estimated maximum electron-hole pair density generated by our pump laser is on the order of $\SI{e18}{\per\cubic\centi\meter}$ (see \textbf{\ztitleref{Methods}}), what is at least an order of magnitude lower than the concentrations necessary for the Auger processes to take place \cite{Wang1990,Borri1997}. Moreover, the Auger recombination is known to be followed by a drop of the PL efficiency with the pumping power density, a behaviour opposite to the clear increase observed in our structure, as presented in \subfigrefL{fig3}{c}. The pump power-dependent PL experiment suggests that the number of the electron-hole pairs in the QW is significantly lower than the estimated values, as the time-integrated PL does not exhibit characteristic intensity saturation even up to $P_{exc}=\SI{e4}{\micro\watt}$ (see \subfigrefL{fig3}{c}). This indicates the likelihood of other non-radiative recombination processes, unrelated to the Auger ones, governing the carrier recombination just after the photoexcitation, and effectively decreasing the measured PL decay times. They could come from carrier relaxation to the states below the bandgap, such as defect states \cite{Michler1994,Hangleiter2019,Zhang2007}.

The $\Gamma$ exciton is characterized by a slightly larger $\tau_{PL}$ of around \SI{19}{\pico\second}. However, its decay deviates from purely monoexponential at early times, with noticeably longer rise time (see the blue curve in \subfigref{fig5}{a}). 
The elongated rise time for the $\Gamma$ exciton is related to the inefficient capture of photo-generated carriers/excitons from the barrier, which may also stem from non-radiative processes happening in this layer, decreasing the overall capture probability. On the contrary, a relatively short capture time for the $X$ valley can be driven by carrier relaxation in the barrier, additionally enhanced by the carrier transfer from the $\varGamma$ to the $X$ valley.  

Furthermore, we studied the temporal evolution of the PL resonance intensity when the temperature is increased and the results can be seen in \subfigref{fig5}{b}, with corresponding time-integrated spectra presented in \subfigref{fig4}{a}. At elevated temperatures, the decays of all resonances become multiexponential, with the shorter component being on the same order of magnitude as the low temperature and the longer component strongly increasing up to several nanoseconds. This makes the net decay much longer, extending above the available \SI{2}{\nano\second} temporal window. Importantly, the $X_{X,Y}$ is characterised by the near resolution-limited decay for all temperatures at which it is observed, which can be explained by the ultrafast transfer of electrons to the true ground state, acting as an efficient nonradiative channel for this state. As the transition probability of the $X_{X,Y}$ state is lower due to the momentum-forbidden nature, the intra-level transitions become more effective. It further visualises the complex dynamics within the system, with possible transitions between closely neighbouring levels, especially when phonons help to conquer the energy barriers and carry particle momentum. Additionally, the temperatures corresponding to the stark elongation of the $X_Z$ and $\Gamma$ emission ($\SIrange{\sim80}{\sim100}{\kelvin}$) match the localization energies estimated from the transition energy evolution as described in the section \textbf{\ztitleref{temp}}, pointing to the likelihood of localization effects playing an additional role in the change of dynamics with temperature, as it has been observed before \cite{Langbein1996}. Even more, the reservoir of trap states existing below the fundamental bandgap of the system, indicated by other measurements, could also contribute to the PL elongation. As the temperature rises, carriers may be released from the trap states and populate the QW confined states due to the increasing phonon bath. This process elongates the observed radiative lifetime \cite{Michler1994,Hangleiter2019,Zhang2007,Rogowicz2021}.

\section*{Summary and Discussion}

In this work, we investigated fundamental optical excitations in the AlGaAs/AlAs QWs near the regime of $\varGamma$-$X$ valley coupling for confined electrons, using the time-integrated and time-resolved photoluminescence. Our experimental characterization is supported by theoretical calculations within the effective-mass framework. 

The low-temperature PL experiments reveal three clearly resolved optical transitions, which we interpreted as direct and momentum- and spatially-indirect excitons. 
The strongest PL feature is attributed to the $\Gamma$ exciton recombination. It has been previously observed in the microcavity-embbedded AlGaAs/AlAs QW \cite{Suchomel2017} and its transition energy matches very well the calculated one. It dominates the spectrum at low temperature and low photo-excitation, as it is expected form a direct exciton transition of a high transition probability.

The other two transitions below the $\Gamma$ exciton have not be thoroughly investigated in similar QW structures. Our study suggests their indirect nature and the $X$-valley-electrons origin. The middle energy transition, pinpointed as $X_{X,Y}$, have features corresponding well to the momentum and spatially indirect exciton, with the $x_{x,y}$-electron constituent. This state is expected to be purely indirect in space and in momentum, hence initially forbidden, yet allowed by a random potential which includes all wavevectors \cite{Finkman1987}. It does not mix with the $\Gamma$ state by the structure potential \cite{Finkman1987} (in contrast to $X_Z$). In our experiments this state is characterized by a largely extended diffusion, and its energy blueshifts with density, as it is expected from dipolar species, affected by repulsive interactions. In both of those observations $X_{X,Y}$'s characteristics clearly differ from the other two states, as one can expect from their momentum-direct (or pseudo-direct) nature. At the highest excitation powers and increased temperatures it becomes unresolvable from the spectra, overwhelmed by the other two transitions. This can also be pinpointed to its indirect nature, as the momentum-selection rules make its recombination the least probable. With increased phonon influence, the $x_{x,y}$ electrons can clearly relax or scatter into the other two states, preventing the $X_{X,Y}$ exciton radiative recombination (with $\Gamma-X_{X,Y}$ mixing allowed by the potential fluctuations due to interface roughness \cite{Feldmann1992}). 

Finally, we have identified the ground state of this system as the $X_{Z}$ exciton, effectively recognizing the whole structure to be of the type-II (spatially indirect) and pseudo-direct in momentum. Symmetry (the structure potential) allows $X_{Z}$'s recombination without the assistance of phonons and its mixing with the $\Gamma$ state  \cite{Finkman1987,Feldmann1992,VanKesteren1989}. Accordingly, in all of our measurements its characteristics are similar to the highest $\Gamma$ state, e.g. it presents similar diffusion profile at low temperatures and redshifts in energy at increased densities, due to the unintentional sample heating. The largest effective mass makes it the lowest energy state in the system  \cite{Finkman1987}. It also makes it prone to the interface roughness and fluctuations and becoming localized. One can clearly see this effect in its observed narrow diffusion profile, the profile's expansion at increased temperatures, in the clear deviation from the Varshni curve at low temperatures, but also in a significant increase of its luminescence and decay time at elevated temperatures, as the excitons that are localized at low temperature become mobile at elevated temperatures  \cite{VanKesteren1989}. The temperature of this increase in the decay constant corresponds well to the estimated localization energy of around \SI{7}{\milli\electronvolt}. 

Our clear identification and characterization of these states are crucial in a careful design of future devices, where specific excitonic states are at the core of interest. Our system offers a superb monolithic platform in which three species of excitons, both direct and indirect, can be exploited. As a clear example, such a QW system can be integrated inside a microcavity, where the dipolar excitons coupled to the cavity field can provide new functionalities and device applications.

\section*{Methods}\zlabel{Methods}

\subsection*{Sample growth}

The investigated structure was grown by molecular beam epitaxy on a nominally undoped (001) GaAs substrate. The structure consists of twelve \SI{9}{\nano\meter}-wide Al$_{0.20}$Ga$_{0.80}$As QWs, separated by \SI{4}{\nano\meter} AlAs barriers, distributed in three stacks of four. The stacks were initially placed in a $\lambda/2$-AlAs cavity surrounded by AlAs/Al$_{0.40}$Ga$_{0.60}$As distributed Bragg reflectors (DBRs), consisting of 28/24 mirror pairs in the bottom/top reflector, including \SI{3}{\nano\meter} GaAs smoothing layers after each mirror pair. Such a full microcavity structure was initially designed for room-temperature polaritonics, however, here we investigate the system prior to the coupling with the photonic modes.  To characterize the QW system, the top DBR has been etched away, so the QW part on a bottom DBR-structure is studied in this work. The resulting structure is characterised by an enhanced luminescence extraction efficiency. Additionally, it was verified that no measurable signatures of remaining photonic modes in the spectral vicinity of the QWs were observed. Thus, one can assume no photonic mode coupling and focus on the QW emission. The structure with a top DBR mirror, providing a full cavity system, has been presented elsewhere \cite{Suchomel2017}.

\subsection*{Optical experiments}

For the time-integrated (PL) and time-resolved photoluminescence (TRPL) experiments, the structure was held in a helium-flow optical cryostat, allowing for the sample temperature control in the range of \SIrange{4.8}{300}{\kelvin}. The structure was excited by \SI{\sim200}{\femto\second}-long laser pulses with the pulse central wavelength of \SI{620}{\nano\meter} ($E_{exc}=$ \SI{2.00}{\electronvolt}). The pulses were delivered by a synchronously-pumped optical parametric oscillator with a \SI{76}{\mega\hertz} repetition frequency, pumped with a Ti:Sapphire laser. The optical excitation was focused on the sample surface via an infinity-corrected high-numerical-aperture (NA) objective (NA=0.65). The PL/TRPL signal was then collected by the same objective and directed to a \SI{0.3}{\meter}-focal length monochromator for spectral resolution. The PL was registered by a thermo-electrically-cooled Si-based electron-multiplied-CCD camera or by a Si-cathode-based streak camera. The TRPL measurement system provided a time resolution of \SI{\sim7}{\pico\second}. Spatial resolution in the time-integrated spatial diffusion measurements was achieved using the two-dimensional CCD camera  \cite{Pieczarka2017,Rudno2018}.

\subsection*{Band structure calculations}

The band structure calculations were performed within the effective mass approximation. Due to the wide band gaps in the investigated materials the conduction band was assumed parabolic near the $\Gamma$ and $X$ points and the spin–orbit split-off (SO) energy is large enough to ignore the SO band. The remaining valence bands (heavy and light hole) were described in the Luttinger model \cite{Luttinger1956}. The energies and wave functions of confined carriers were obtained numerically by solving the Schr\"odinger equation using the method reported in Ref. \cite{Kubisa2003}. Standard material parameters \cite{Piperek2003} for GaAs and AlAs were used in the calculation and linear interpolation was applied for Al$_{0.20}$Ga$_{0.80}$As. The valence-band offset was calculated as $37\%$ of the difference in band gaps in the well and barrier layers. The exciton binding energy was estimated from the formula $E_{Ry}= \mu e^4/2\epsilon^2\hbar^2$, where $\mu$ is the in-plane reduced effective mass and $\epsilon$ is the dielectric constant.

\subsection*{Photo-excited electron-hole pair density estimation}
The estimation of the maximum electron-hole pair density $N_{e-h}$, generated by our pump laser and used in sections  \textbf{\ztitleref{excitation}} and \textbf{Time-resolved photoluminescence} has been done with a following formula: 
\begin{equation}
    N_{e-h}=\frac{P_{exc}}{A f E_{exc}} T_{obj} T_{glass} (1-R_{sample}) e^{-\alpha d},
\end{equation}
where  $P_{exc}$ is the measured averaged excitation power, $A=\pi \left( \phi/2 \right)^2$ is the laser spot area, $\phi=$ \SI{4}{\micro\meter} is its diameter, $f= \SI{76}{\mega\hertz}$ is the laser pulse repetition frequency, the excitation energy $E_{exc}=\SI{2.00}{\electronvolt}=\SI{3.2038e-19}{\joule}$, $T_{obj}=80~\%$ is the objective transmission at \SI{620}{\nano\meter}, the sample reflection under the normal incidence $R_{sample}=30~\%$\cite{Aspnes1986} and the cryostat glass transmission is $T_{glass}=96~\%$. $\alpha=\SI{2e4}{\per\centi\meter}$ is the absorption coefficient for the Al$_{0.2}$Ga$_{0.8}$As QW material at cryogenic temperatures\cite{Monemar1976} and $d$ is the effective width of the absorption layer (QW stack), $d=4\cdot3\cdot9$\ nm=$\SI{108}{\nano\meter}$. The experiment with $P_{exc}=\SI{100}{\micro\watt}$ gives the maximum concentration of $N_{e-h}=\SI{1.41e13}{\per\square\centi\meter}$ ($N_{e-h}/d=\SI{1.31e18}{\per\cubic\centi\meter}$).
\\
\section*{Data availability}
The datasets generated during and/or analysed during the current study are available from the corresponding author on reasonable request.

\bibliography{bibliography}

\section*{Acknowledgements}
D. B. and M. S. acknowledge financial support from the National Science Centre Poland within the (Grant No. 2018/30/E/ST7/00648). 
The W\"{u}rzburg group acknowledges financial support by the German Research Foundation (DFG) under Germany’s Excellence Strategy–EXC2147 “ct.qmat” (project id 390858490).

\section*{Author contributions statement}
D. B. conducted all the spectroscopic experiments and analysed the experimental data, K.R. and M. K. performed the theoretical calculations, S. K., S. H and C. S provided the QW structure, D. B., M. P., S.K., C. S, S. H., and M.S. analysed and discussed the results. D.B. wrote the first version of the manuscript and prepared all figures. All authors reviewed the manuscript to its final form. 

\section*{Additional information}
The authors declare no competing interests.

\end{document}